\documentclass[aps,prl,reprint,superscriptaddress,longbibliography,floatfix]{revtex4-2}

\usepackage{amsmath}
\usepackage{amssymb}
\usepackage{graphicx}
\usepackage{placeins}
\usepackage{longtable}
\usepackage{array}
\usepackage{bm}
\usepackage{xcolor}
\usepackage{hyperref}
\hypersetup{colorlinks=true, linkcolor=blue, citecolor=blue, urlcolor=blue}
\makeatletter
\long\def\@makecaption#1#2{%
  \par\vskip\abovecaptionskip
  \begingroup
    \small\rmfamily\samepage\flushing
    \let\footnote\@footnotemark@gobble
    \@make@capt@title{#1}{#2}\par
  \endgroup
  \vskip\belowcaptionskip
}
\newenvironment{suppfigure}{%
  \par\addvspace{5pt}\noindent\begin{minipage}{\linewidth}%
  \def\@captype{figure}\centering
}{\end{minipage}\par\addvspace{5pt}}
\newenvironment{supptable}{%
  \par\addvspace{5pt}\noindent\begin{minipage}{\linewidth}%
  \def\@captype{table}\centering
}{\end{minipage}\par\addvspace{5pt}}
\makeatother

\newcommand{\papertitle}{Scalable Simulation of Strongly Correlated Fermions\texorpdfstring{\\}{ }with Neural Gaussian Tensor Networks}

\newcommand{\Pf}{\operatorname{Pf}}
\newcommand{\ket}[1]{|#1\rangle}
\newcommand{\bra}[1]{\langle#1|}
\newcommand{\braket}[2]{\langle#1|#2\rangle}

\colorlet{ampgray}{black!50}

\newcounter{algline}
\newenvironment{alglines}{\begin{list}{\footnotesize\arabic{algline}:}{\usecounter{algline}%
    \setlength{\leftmargin}{2.2em}\setlength{\labelwidth}{1.8em}\setlength{\itemsep}{0pt}%
    \setlength{\parsep}{0pt}\setlength{\topsep}{2pt}}}{\end{list}}
\newcommand{\algind}[1]{\hspace*{#1\dimexpr 1.5em\relax}}

\newcounter{algorithm}
\newenvironment{algorithm}[1][]{%
    \par\addvspace{\intextsep}\noindent
    \minipage{\linewidth}%
    \def\caption##1{%
        \refstepcounter{algorithm}%
        \hrule height 0.8pt \nobreak \vskip 3pt
        {\flushing \textbf{Algorithm~\thealgorithm.} ##1\par}%
        \nobreak \vskip 2pt \hrule height 0.4pt \nobreak \vskip 3pt
    }%
}{%
    \vskip 3pt \hrule height 0.8pt
    \endminipage\par\addvspace{\intextsep}%
}

\begin{document}
\setcounter{dbltopnumber}{3}
\renewcommand{\dbltopfraction}{0.95}
\title{\papertitle}
\author{Runze Chi}
\email{mr.rzchi@gmail.com}
\affiliation{Division of Chemistry and Chemical Engineering, California Institute of Technology, Pasadena, California 91125, USA}
\affiliation{Marcus Center for Theoretical Chemistry, California Institute of Technology, Pasadena, California 91125, USA}
\author{Ao Chen}
\affiliation{Division of Chemistry and Chemical Engineering, California Institute of Technology, Pasadena, California 91125, USA}
\affiliation{Marcus Center for Theoretical Chemistry, California Institute of Technology, Pasadena, California 91125, USA}
\author{Garnet Kin-Lic Chan}
\email{gkc1000@gmail.com}
\affiliation{Division of Chemistry and Chemical Engineering, California Institute of Technology, Pasadena, California 91125, USA}
\affiliation{Marcus Center for Theoretical Chemistry, California Institute of Technology, Pasadena, California 91125, USA}
\date{}

\begin{abstract}
Accurate simulations of strongly correlated electrons remain a central challenge in computational physics. We introduce neural-network Gaussian projected entangled-pair states (NN-GPEPS), which exploit the locally-entangled subspace of Gaussian states to give an architecture more compatible with modern neural networks and an exact contraction cost of $O(N^{3/2})$ on two-dimensional lattices. For the Hubbard model at $U/t=8$ and $1/8$ hole doping, we obtain state-of-the-art variational energies in large lattices of up to 1024 sites. These results demonstrate an accurate approach to strongly correlated fermions by connecting neural networks with Gaussian tensor networks.
\end{abstract}
\maketitle

Devising compact representations of strongly correlated fermionic states is a central challenge in quantum many-body physics. In recent years, neural networks have been explored as flexible variational representations of wavefunctions~\cite{Carleo_Science17_NQS}. To capture the sign structure associated with fermions, it is necessary to incorporate an underlying bias into the neural network ansatz. Some widely used approaches for this thus use the neural network to parametrize  determinant or Pfaffian (Gaussian) wavefunctions~\cite{Luo_PRL19_Backflow,Moreno_PNAS22_HFDS,Chen_PNAS26_HFPS,Rende_arxiv26_SBP}, an ansatz often known as neural network backflow (NNBF).
Fermionic tensor network states (TNS) provide an alternative representation of fermionic signs, incorporating them through a $\mathbb{Z}_2$-graded tensor algebra~\cite{Corboz_PRB10_fPEPS,Kraus_PRA10_fPEPS,Mortier_SP25_fTN,Liu_PRL25_Hubbard-fPEPS}. Recent work has demonstrated the effectiveness of neural-network parametrized TNS (NN-TNS)~\cite{Du_PRB26_NN-fTNS}. Both NNBF and NN-TNS have achieved highly competitive variational energies in challenging fermionic lattice problems~\cite{Gu_PRX-Intelligence_SCALE,Rende_arxiv26_SBP,Du_PRB26_NN-fTNS}.

\begin{figure*}[tp]
\centering
\includegraphics[width=\textwidth]{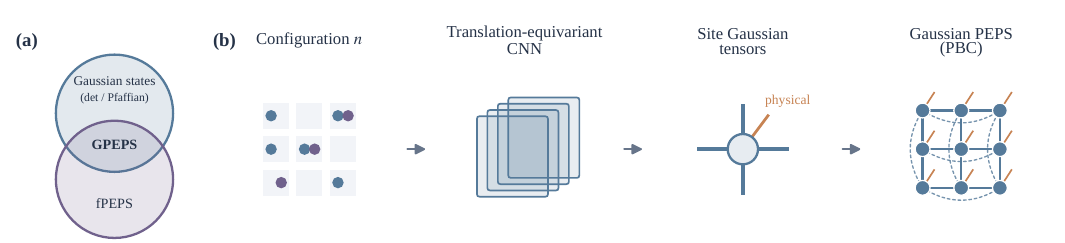}
\setlength{\abovecaptionskip}{4pt}
\caption{Construction of NN-GPEPS. (a) GPEPS as an intersection of Gaussian states and fPEPS. (b) A translation-equivariant convolutional neural network (CNN) with circular padding maps an occupation configuration $n$ to local pairing matrices $Z_s(n)$, which define Gaussian states on the physical and virtual modes at each site. Each site has two physical modes and $N_v$ virtual modes on each of its four legs. Contracting neighboring virtual modes with fixed entangled bond states gives the Gaussian PEPS $|G_\theta(n)\rangle$. Projection onto $|n\rangle$ yields the amplitude $\psi_\theta(n)$. Dashed bonds denote periodic boundaries.}
\label{fig:structure}
\end{figure*}

A general Gaussian state requires $O(N^2)$ pairing-matrix elements. However, modern neural architectures such as convolutional neural networks (CNNs) and transformers naturally produce sitewise features, $O(N)$ in total. Tensor network states, in contrast, provide a local parametrization compatible with $O(N)$ features, but their evaluation on large two-dimensional lattices typically relies on approximate contraction. Gaussian tensor network states (GTNS) lie at the intersection of these two families~\cite{Bravyi_QIC05_Lagrangian,Schuch_PRB19_GaussianMPS,Yang_PRB23_ProjectedDWave}, as illustrated in Figure~\ref{fig:structure}(a). Built from local Gaussian tensors, they describe Gaussian states with area-law entanglement. In the current context, the essential feature is that they combine an $O(N)$ local parametrization with the possibility of efficient \emph{exact} contraction.

Utilizing the advantages of GTNS, we introduce neural-network Gaussian projected entangled-pair states (NN-GPEPS). Analogous to NN-TNS, neural networks in NN-GPEPS generate configuration-dependent local Gaussian tensors [Fig.~\ref{fig:structure}(b)]. This yields a generally non-Gaussian wavefunction, while preserving the efficient exact contraction structure of GTNS for each amplitude. The ansatz can be systematically improved by increasing the number of virtual fermionic modes as well as the expressivity of the neural network. The amplitude can be exactly computed with $O(N^{3/2})$ cost~\cite{Thomas_PRE13_Pfaffian} in two dimensions, and generally $O(N^{3-3/p})$ in $p \geq 2$ dimensions. If truncation is allowed, approximate Gaussian boundary contraction can further reduce the cost to $O(N)$~\cite{Schuch_PRB19_GaussianMPS}.
We apply the NN-GPEPS ansatz to study the Hubbard model at $U/t=8$ and $1/8$ hole doping on the $16\times4$ and $16\times16$ lattices. There, we achieve results that are competitive with, or improve upon, all prior results in the literature.
Further, we show that direct transfer of the optimized parameters to a $32\times32$ lattice also achieves, without further optimization, state-of-the-art variational energies on the larger lattice.

\textit{Method.---}We consider the repulsive Hubbard model on a periodic lattice of $N$ sites,
\begin{equation}
H=-t\sum_{\langle ij\rangle,\sigma}(c^\dagger_{i\sigma}c_{j\sigma}+\mathrm{h.c.})
+U\sum_i n_{i\uparrow}n_{i\downarrow},
\label{eq:hubbard}
\end{equation}
where $c^\dagger_{i\sigma}$ creates an electron with spin $\sigma$, $n_{i\sigma}=c^\dagger_{i\sigma}c_{i\sigma}$, and $\langle ij\rangle$ counts each nearest-neighbor bond once. In this paper, we set $t=1$ and focus on $U=8$ at $1/8$ hole doping.

To approximate its ground state, we introduce a NN-GPEPS ansatz, which takes the form
\begin{equation}
\psi_\theta(n) = \langle n|G_\theta(n)\rangle,
\label{eq:nn-gpeps}
\end{equation}
where $|n\rangle$ is an occupation-basis state, and $\theta$ represents parameters in the neural network. For each configuration $n$, we construct the Gaussian PEPS by contracting the virtual modes of local Gaussian states~\cite{Kraus_PRA10_fPEPS},
\begin{equation}
|G_\theta(n)\rangle=(\mathbf{1}_{\mathrm{phys}}\otimes\langle\Omega|)
\bigotimes_s |Z_s(n)\rangle,
\label{eq:gaussian-network}
\end{equation}
where $|Z_s(n)\rangle$ is a local fermionic Gaussian state containing two physical modes and $N_v$ virtual modes on each of its four legs. It is specified by $m(m-1)/2$ independent real parameters in a skew-symmetric pairing matrix $Z_s(n)$, with $m=2+4N_v$. The matrix entries specify pair creation amplitudes among the modes at site $s$. The identity operator $\mathbf{1}_{\mathrm{phys}}$ acts on the physical modes (leaving them uncontracted), while $\langle\Omega|$ contracts neighboring virtual modes through normalized entangled bond states $(|00\rangle+|11\rangle)/\sqrt{2}$. We generate the independent entries of $Z_s(n)$ using a translation-equivariant CNN with trainable parameters $\theta$ and circular padding, which implements the periodic boundary conditions [Fig.~\ref{fig:structure}]. Combined with uniform virtual bonds, this construction encodes lattice translation symmetry directly in the ansatz. The network architecture is given in Appendix~\ref{app:network}.
The full wavefunction $|\Psi_\theta\rangle$ is generally non-Gaussian because a different set of local pairing matrices is generated for each configuration.

For each amplitude $\psi_\theta(n)$, we evaluate the overlap in Eq.~\eqref{eq:nn-gpeps} by contracting the virtual bonds and projecting the physical modes onto $n$. Using the Gaussian overlap identities in Appendix~\ref{app:contraction}, we obtain
\begin{equation}
\begin{aligned}
M_\theta(n)&=\tilde n\star\left[\bigoplus_s Z_s(n)+B\right]\star\tilde n,\\
\psi_\theta(n)&=2^{-b/2}\Pf M_\theta(n).
\end{aligned}
\label{eq:pfaffian}
\end{equation}
Here $\bigoplus_s Z_s(n)$ describes the local Gaussian states, while $B$ is a fixed sparse skew-symmetric matrix representing the virtual bonds. The selection $\tilde n$ retains all virtual modes and the occupied physical modes of $n$ and $\star$ slices the corresponding rows and columns. The number of virtual mode pairs is $b=2NN_v$. Note that, although $M_\theta(n)$ is sparse in the expanded space of physical and virtual modes, the amplitude is the Pfaffian of a dense matrix on the physical modes: labeling the virtual and occupied physical modes $V$ and $P$, $\psi_\theta(n) = 2^{-b/2} \Pf(M_{VV}) \Pf(M_{PP} - M_{PV} M_{VV}^{-1} M_{VP})$.

Nested-dissection Pfaffian elimination exploits the two-dimensional sparsity of $M_\theta(n)$ to give the exact contraction cost
\begin{equation}
T_{\rm TN}=O(N_v^3N^{3/2}).
\label{eq:scaling}
\end{equation}
The computational cost of the neural network scales as $O(N)$, and the contraction cost can also be reduced to $O(N)$ by using approximate boundary contraction~\cite{Schuch_PRB19_GaussianMPS}.

We optimize the neural-network parameters $\theta$ by minimizing the variational energy using variational Monte Carlo (VMC)~\cite{Becca_17_VMCtext}.
The automatic differentiation provides the derivatives needed for optimization. We use these derivatives to optimize the parameters with the AdamSR optimizer~\cite{quantax} and the iterative least-squares minimal residual (LSMR) solver~\cite{Fong_SIAM11_LSMR}.
After optimizing the original wavefunction, we impose mirror reflections in the $x$ and $y$ directions and spin inversion through a projection over eight symmetry-group elements~\cite{Nomura_JPCM21_restoration,Chen_PNAS26_HFPS}. The symmetry-projected wavefunction is then further optimized.
During training, CNN products use TensorFloat-32 (TF32), while Gaussian tensor network contraction uses single (FP32) or double (FP64) precision. For final-state measurements, we use FP32 for both CNN products and Gaussian tensor network contraction. Optimization and measurement protocols are given in Appendix~\ref{app:recipe}.

\textit{Why NN-GPEPS?---}
It seems that NN-GPEPS is simply another way to encode fermionic sign structures in neural wavefunctions, while here we want to explain why it is a much more natural choice compared with NNBF or NN-fPEPS.

In NNBF, the Gaussian state is essentially a long-range correlated state, requiring $\mathcal{O}(N^2)$ information to construct the correlation matrix. Nevertheless, the information provided by the output of modern neural networks, either convolutional neural networks (CNNs) or transformers, represents local properties at each site and scales as $\mathcal{O}(N)$. The mismatch between $\mathcal{O}(N^2)$ and $\mathcal{O}(N)$ becomes an unfavorable scaling in large systems. It forces one to use $\mathcal{O}(N)$ neural outputs as a sparse representation of $\mathcal{O}(N^2)$ elements in Gaussian states, transformed through simple linear expansions without a clear physical interpretation~\cite{Chen_PNAS26_HFPS, Gu_NC26_NQS-Hubbard, Gu_PRX-Intelligence_SCALE, Rende_arxiv26_SBP}. In GPEPS, we instead choose the area-law entangled subspace of Gaussian states that requires only local parametrization with $\mathcal{O}(N)$ elements from neural outputs. This is a much more natural choice than a normal Gaussian state. The $\mathcal{O}(N^{3/2})$ computational cost of GPEPS is also lower than the $\mathcal{O}(N^3)$ cost of Gaussian states.

On the other hand, NN-fPEPS requires a favorable amount of $\mathcal{O}(N)$ information due to area-law entanglement, and it indeed shows great accuracy in small systems where exact contractions are available. Nevertheless, exact contraction of general PEPS in two or more dimensions is exponentially costly. Although approximate contractions with $\mathcal{O}(N)$ complexity are available, we observe a clear loss of accuracy when these approximate methods are employed. GPEPS, instead, allows exact contractions with $\mathcal{O}(N^{3/2})$ cost due to its Gaussian structure, avoiding the problem of approximate contractions.

Therefore, GPEPS provides favorable $\mathcal{O}(N)$ information scaling as well as efficient exact contraction, at the cost of reduced expressivity, which can be compensated for by the expressive power of neural networks. NN-GPEPS hence serves as a natural choice to encode fermionic sign structures in fermionic neural wavefunctions.

\begin{figure}[tbp]
\centering
\includegraphics[width=0.9\columnwidth]{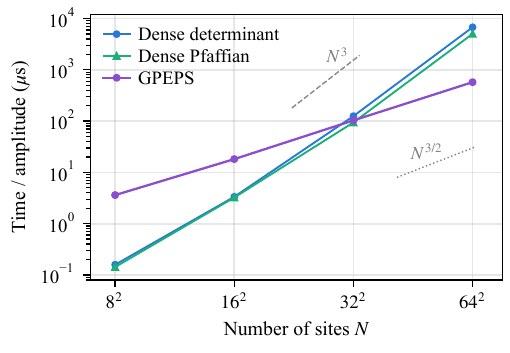}
\caption{Time per amplitude for Gaussian PEPS contraction and dense determinant and Pfaffian evaluation on one A100 80GB GPU, for periodic $L\times L$ lattices with $N=L^2$ sites. GPEPS uses $N_v=3$ and FP32 contraction. Dense evaluations use fermix~\cite{Chen_github26_fermix} with TF32$\times3$ arithmetic. Dashed and dotted lines indicate $N^3$ and $N^{3/2}$, respectively.}
\label{fig:scaling}
\end{figure}

\textit{Amplitude cost.---}We first benchmark the contraction of Gaussian tensor networks on periodic $L\times L$ lattices, with $N=L^2$. We take $N_v=3$, giving three virtual fermionic modes on each leg, and evaluate amplitudes for half-filled configurations using Gaussian tensor network contraction on one A100 80GB GPU. Figure~\ref{fig:scaling} compares FP32 Gaussian tensor network contraction with dense determinant and Pfaffian evaluation using fermix~\cite{Chen_github26_fermix}.

Nested-dissection Pfaffian elimination has a computational cost of $O(N^{3/2})$ at fixed $N_v$, compared with $O(N^3)$ for a dense determinant or Pfaffian. The guiding lines in Fig.~\ref{fig:scaling} show these asymptotic scalings. At $32\times32$, Gaussian tensor network contraction has a cost comparable to dense Pfaffian and determinant evaluation, and at $64\times64$ it is roughly an order of magnitude faster than dense Pfaffian and determinant evaluation.  The determinant and Pfaffian computations in fermix are highly optimized through customized kernels, while our Gaussian tensor network contraction still has room for further optimization. Thus we can expect the time cost of GPEPS to be reduced in the future. Appendix~\ref{app:benchmark} further separates the CNN and contraction costs in a complete amplitude evaluation and compares FP32 and FP64 Gaussian tensor network contraction times.

\textit{Accuracy.---}We first benchmark NN-GPEPS on the $16\times4$ lattice at $U/t=8$ and $1/8$ hole doping. {We note that this small system size is not expected to show the advantage of the GPEPS Gaussian parametrization, but we study it since it is one where existing backflow parametrizations can work well.}
The symmetry-projected $N_v=3$ and $4$ states reach energies per site of $-0.765883(26)$ and $-0.766152(16)$, respectively. The $N_v=4$ energy is competitive with previous NNBF and ACE results~\cite{Loehr_arxiv25_NNBF,Gu_PRX-Intelligence_SCALE}. A joint zero-variance extrapolation of the two projected trajectories gives $-0.7673(5)$. The optimized states also exhibit pronounced spin-stripe correlations and weak finite-distance pairing. Detailed energy comparisons and correlation functions are presented in Appendix~\ref{app:energy-benchmarks}, while the extrapolation procedure is described in Appendix~\ref{app:extrapolation}.

We next assess the accuracy on the larger $16\times16$ square lattice, {where the advantages of the GPEPS can be expected to appear}. For each $N_v$, we reuse the CNN parameters learned on $16\times4$ and continue optimization with the same architecture. The projected $N_v=3$ and $4$ states reach $E/N=-0.759835(14)$ and $-0.760115(11)$, respectively. Both energies improve on the earlier variational energies reported in Table~\ref{tab:energy-comparison}. The $N_v=4$ variational energy is already consistent, within the reported uncertainty, with the MR-nVAFQMC extrapolation~\cite{Shang_arxiv26_MRnVAFQMC}.
A joint zero-variance extrapolation of the projected $N_v=3$ and $4$ optimization trajectories gives $E_0/N=-0.7622(7)$.

\begin{table}[!htbp]
\centering
\caption{Energies and local-energy variances per site at $U/t=8$ and 1/8 hole doping on the periodic $16\times16$ lattice. Parentheses give reported uncertainties. The NN-GPEPS extrapolation procedure is described in Appendix~\ref{app:extrapolation}. A dash denotes an unavailable variance.}
\label{tab:energy-comparison}
\footnotesize\setlength{\tabcolsep}{3pt}
\begin{ruledtabular}
\begin{tabular}{lcc}
Method & $E/N$ & $\sigma^2/N$ \\
\colrule
Transformer backflow~\cite{Gu_NC26_NQS-Hubbard} & $-0.7563$ & --- \\
HFPS~\cite{Roth_arxiv25_HubbardSC} & $-0.7515(1)$ & $0.0638$ \\
SCALE~\cite{Gu_PRX-Intelligence_SCALE} & $-0.7529$ & --- \\
ACE~\cite{Gu_PRX-Intelligence_SCALE} & $-0.7573$ & --- \\
VAFQMC~\cite{Wu_Science24_Vscore,Sorella_PRB23_VAFQMC} & $-0.75865(3)$ & $0.0059(8)$ \\
MR-nVAFQMC~\cite{Shang_arxiv26_MRnVAFQMC} & $-0.7589(19)$ & --- \\
NN-GPEPS, $N_v=3$ & $-0.759835(14)$ & $0.0149(7)$ \\
NN-GPEPS, $N_v=4$ & $-0.760115(11)$ & $0.01287(24)$ \\
MR-nVAFQMC (extrapolated)~\cite{Shang_arxiv26_MRnVAFQMC} & $-0.7604(32)$ & --- \\
NN-GPEPS (extrapolated) & $-0.7622(7)$ & --- \\
\end{tabular}
\end{ruledtabular}
\end{table}

{
To characterize the optimized $16\times16$ states, we measure spin, charge and singlet-pair correlations. We define $C_s(\mathbf r)=\langle\mathbf S_0\cdot\mathbf S_{\mathbf r}\rangle$ and $C_c(\mathbf r)=\langle\delta n_0\delta n_{\mathbf r}\rangle$, where $\delta n_i=n_i-7/8$. The structure factors are $S(\mathbf k)=\sum_{\mathbf r}e^{i\mathbf k\cdot\mathbf r}C_s(\mathbf r)$ and $N_c(\mathbf k)=\sum_{\mathbf r}e^{i\mathbf k\cdot\mathbf r}C_c(\mathbf r)$, and the figures show these divided by $N$. For pairing, we use the normalized nearest-neighbor singlet $B^\dagger_{ij}=(c^\dagger_{i\uparrow}c^\dagger_{j\downarrow}-c^\dagger_{i\downarrow}c^\dagger_{j\uparrow})/\sqrt2$ to define
\begin{equation}
\begin{aligned}
\Delta_i^\dagger&=\tfrac14\sum_{\delta=\pm\hat x,\pm\hat y}f_\delta B^\dagger_{i,i+\delta},\\
C_p(\mathbf r)&=\langle\Delta_0^\dagger\Delta_{\mathbf r}\rangle-\mathcal N_{0,\mathbf r},
\end{aligned}
\label{eq:pair-correlations}
\end{equation}
where $f_{\pm\hat x}=1$ and $f_{\pm\hat y}=-1$. The term $\mathcal N_{0,\mathbf r}$ removes the disconnected products $\langle c^\dagger c\rangle\langle c^\dagger c\rangle$ in the Wick decomposition of $\langle\Delta_0^\dagger\Delta_{\mathbf r}\rangle$.
We define $\Delta_{\rm SC}^2$ as the mean of $C_p(\mathbf r)$ over $d(\mathbf r)\ge d_{\max}/2$, where $d(\mathbf r)$ is the shortest Euclidean distance under periodic boundary conditions and $d_{\max}$ is its maximum over all lattice displacements.
\par
}

\begin{figure}[!tbp]
\centering
\includegraphics[width=\columnwidth]{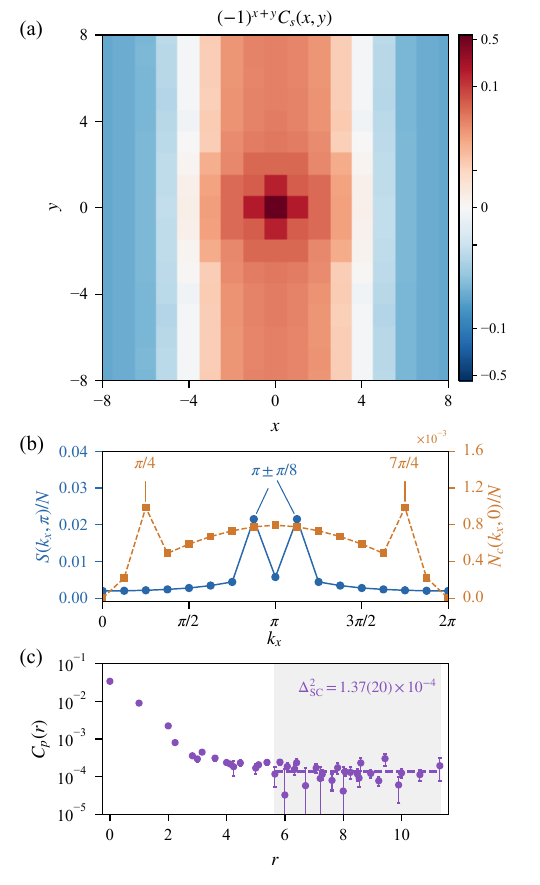}
\setlength{\abovecaptionskip}{4pt}
\caption{Correlations of the projected $16\times16$, $N_v=4$ NN-GPEPS state at $U/t=8$ and $1/8$ doping. (a) Staggered real-space spin correlations. (b) Spin (blue, left axis) and charge (orange, right axis) structure-factor cuts, with peak positions labeled. (c) Connected pair correlations averaged over displacements of equal shortest periodic distance $r$. The gray region marks $r\ge4\sqrt2$, and the dashed line gives the mean over these displacements, $\Delta_{\rm SC}^2$. Panel (a) uses a symmetric logarithmic color scale. Panel (c) shows positive means on a logarithmic axis, while the dashed-line average includes both signs. Error bars show statistical uncertainties.}
\label{fig:orders}
\end{figure}

Figure~\ref{fig:orders} resolves the stripe pattern in the projected $N_v=4$ state. The real-space spin correlations have a modulated antiferromagnetic envelope, with peaks in $S(k_x,\pi)$ at $k_x=\pi\pm\pi/8$ and in $N_c(k_x,0)$ at $k_x=\pm\pi/4$. These spin and charge periods of 16 and 8 are consistent with the filled stripes found in previous studies~\cite{Zheng_Science17_Hubbard,Roth_arxiv25_HubbardSC,Shang_arxiv26_MRnVAFQMC}. Averaging the spin structure factor over $\mathcal Q=\{(\pi\pm\pi/8,\pi),(\pi,\pi\pm\pi/8)\}$ gives $M_{\rm str}^2=\frac{1}{4N}\sum_{\mathbf Q\in\mathcal Q}S(\mathbf Q)=0.013725(10)$. The stripe strength is comparable to that found with MR-nVAFQMC~\cite{Shang_arxiv26_MRnVAFQMC}.

Pair correlations decrease rapidly with distance but retain a small positive mean, $\Delta_{\rm SC}^2=1.37(20)\times10^{-4}$, over $r\ge4\sqrt2$. MR-nVAFQMC finds the same distant-window measure compatible with zero~\cite{Shang_arxiv26_MRnVAFQMC}, while HFPS finds long-distance pairing correlations nearly zero within statistical uncertainties~\cite{Roth_arxiv25_HubbardSC}. The resolved finite-size signal here does not establish superconducting long-range order, which requires scaling to larger lattices. The $N_v=3$ and $4$ states show similar qualitative stripe and pairing behavior on each lattice (Appendix~\ref{app:observable-convergence}).

\textit{Parameter transfer to the $32\times32$ lattice.---}Finally, we test a direct transfer from the $16\times16$ to $32\times32$ lattice using the unprojected $N_v=3$ and $4$ states. We reuse the learned CNN parameters without further optimization and evaluate each state without and with the symmetry projection. Both sampling and evaluation use FP32 (Appendix~\ref{app:precision}). The transferred states give energies per site of $-0.757451(21)$ and $-0.758019(22)$, respectively. Applying the projection at evaluation lowers the energies to $-0.758414(27)$ and $-0.758727(33)$, with variances per site of $0.01647(64)$ and $0.01606(82)$. All four energies lie substantially below the SCALE and SCALE+GFMC values and are comparable to the MR-nVAFQMC (including extrapolation) results, within their large uncertainty~\cite{Gu_PRX-Intelligence_SCALE,Shang_arxiv26_MRnVAFQMC}. This improvement is obtained without enlarging the network or performing an optimization at the target size. The projected transferred states retain variance densities comparable to those on $16\times16$, providing a useful starting point for further optimization.

\begin{table}[!tbp]
\centering
\caption{Energies and local-energy variances per site on $32\times32$ at the parameters of Table~\ref{tab:energy-comparison}. NN-GPEPS results are obtained by direct transfer of unprojected $16\times16$ states, evaluated without and with (P) the symmetry projection. Other rows give published finite accuracy results and an estimated extrapolation to the exact energy. A dash denotes an unavailable variance.}
\label{tab:reuse}
\footnotesize\setlength{\tabcolsep}{3pt}
\begin{ruledtabular}
\begin{tabular}{lcc}
Method & $E/N$ & $\sigma^2/N$ \\
\colrule
NN-GPEPS $N_v=3$ & $-0.757451(21)$ & $0.02407(51)$ \\
NN-GPEPS $N_v=4$ & $-0.758019(22)$ & $0.02013(42)$ \\
NN-GPEPS $N_v=3$ (P) & $-0.758414(27)$ & $0.01647(64)$ \\
NN-GPEPS $N_v=4$ (P) & $-0.758727(33)$ & $0.01606(82)$ \\
\colrule
SCALE~\cite{Gu_PRX-Intelligence_SCALE} & $-0.7501$ & --- \\
SCALE + GFMC~\cite{Gu_PRX-Intelligence_SCALE} & $-0.7548$ & --- \\
MR-nVAFQMC~\cite{Shang_arxiv26_MRnVAFQMC} & $-0.7577(26)$ & --- \\
MR-nVAFQMC (extrapolated)~\cite{Shang_arxiv26_MRnVAFQMC} & $-0.7581(17)$ & --- \\
\end{tabular}
\end{ruledtabular}
\end{table}

\textit{Conclusion and discussion.---}We introduce neural-network Gaussian projected entangled-pair states for strongly correlated fermions. The Gaussian tensor-network structure enables exact amplitude evaluation with $O(N^{3/2})$ cost in two dimensions, while neural networks make the local tensors depend on the occupation configuration. For the doped Hubbard model, NN-GPEPS achieves state-of-the-art variational energies on $16\times16$, while direct parameter transfer yields competitive energies on $32\times32$ without further optimization.

The observed accuracy suggests that the local Gaussian parametrization provides an architecture more compatible with neural networks, while retaining efficient exact contraction. These results highlight the close connection between tensor networks and neural-network wavefunctions and provide a practical route to scalable studies of strongly correlated fermions and their phase diagrams.

\section*{Acknowledgments}
\begin{acknowledgments}
We thank Si-Jing Du and Gunhee Park for helpful discussions.
This work was supported by the U.S. Department of Energy, Office of Science, Accelerated Research in Quantum Computing Centers, through the Quantum Utility through Advanced Computational Quantum Algorithms (QuACQ) program under Award No.~DE-SC0025572.
The calculations were carried out mainly at the National Energy Research Scientific Computing Center (NERSC), a U.S. Department of Energy Office of Science User Facility at Lawrence Berkeley National Laboratory operated under Contract No.~DE-AC02-05CH11231, using NERSC awards \mbox{ASCR-ERCAP0037872} and \mbox{DDR-ERCAP0038682}, and in part at the Caltech Resnick High Performance Computing Center.
We used the AI coding assistants Claude Code (Anthropic) and Codex (OpenAI) to help develop the simulation and analysis code, launch and monitor jobs on the computing clusters, prepare figures, and draft parts of the manuscript.
The authors reviewed all AI-assisted code, results and writing and are responsible for the final content.
\end{acknowledgments}

\section*{Data and code availability}
The code and data supporting this work are available at Ref.~\cite{Chi_github26_neuralGPEPS}.

\bibliography{reference}

\clearpage
\onecolumngrid
\makeatletter
\let\supp@startsection\@startsection
\def\@startsection#1#2#3#4#5#6{%
  \supp@startsection{#1}{#2}{#3}{12pt plus 2pt minus 2pt}{6pt}{#6}}
\makeatother
\setlength{\abovedisplayskip}{6pt plus 2pt minus 2pt}
\setlength{\belowdisplayskip}{6pt plus 2pt minus 2pt}
\setlength{\abovedisplayshortskip}{3pt plus 2pt}
\setlength{\belowdisplayshortskip}{4pt plus 2pt minus 2pt}
\setlength{\textfloatsep}{12pt plus 2pt minus 2pt}
\setlength{\floatsep}{10pt plus 2pt minus 2pt}
\setlength{\intextsep}{10pt plus 2pt minus 2pt}
\setcounter{secnumdepth}{2}
\appendix
\numberwithin{figure}{section}
\numberwithin{table}{section}
\setcounter{figure}{0}
\setcounter{table}{0}
\renewcommand{\thefigure}{\thesection\arabic{figure}}
\renewcommand{\thetable}{\thesection\arabic{table}}
\renewcommand{\theHfigure}{appendix.\thesection.\arabic{figure}}
\renewcommand{\theHtable}{appendix.\thesection.\arabic{table}}
\numberwithin{algorithm}{section}
\setcounter{algorithm}{0}
\renewcommand{\thealgorithm}{\thesection\arabic{algorithm}}
\section{Gaussian tensor network contraction}
\label{app:contraction}

An even fermionic Gaussian state in pairing form is
\begin{equation}
    \ket{\Psi_Z} = \exp\left(\sum_{a<b} Z_{ab}\, c^\dagger_a c^\dagger_b \right) \ket{0}
    = \prod_{a<b} (1 + Z_{ab}c^\dagger_a c^\dagger_b) \ket{0}
    \label{eq:gauss}
\end{equation}
where the summation covers all $m$ fermionic modes, and $Z$ is a skew-symmetric $m \times m$ matrix.
The state is unnormalized, with vacuum amplitude $\braket{0}{\Psi_Z}=1$.
For an input basis state $n_I$, its wavefunction amplitude is
\begin{equation}
    \psi_Z(n_I) = \braket{n_I}{\Psi_Z} = \Pf(n_I \star Z \star n_I),
    \label{eq:minor}
\end{equation}
where $n_I \star Z$ and $Z \star n_I$ slice the rows and columns of $Z$ according to the occupied fermionic modes of $n_I$, respectively.

\begin{figure}[!htbp]
    \centering
    \includegraphics[width=0.80\textwidth]{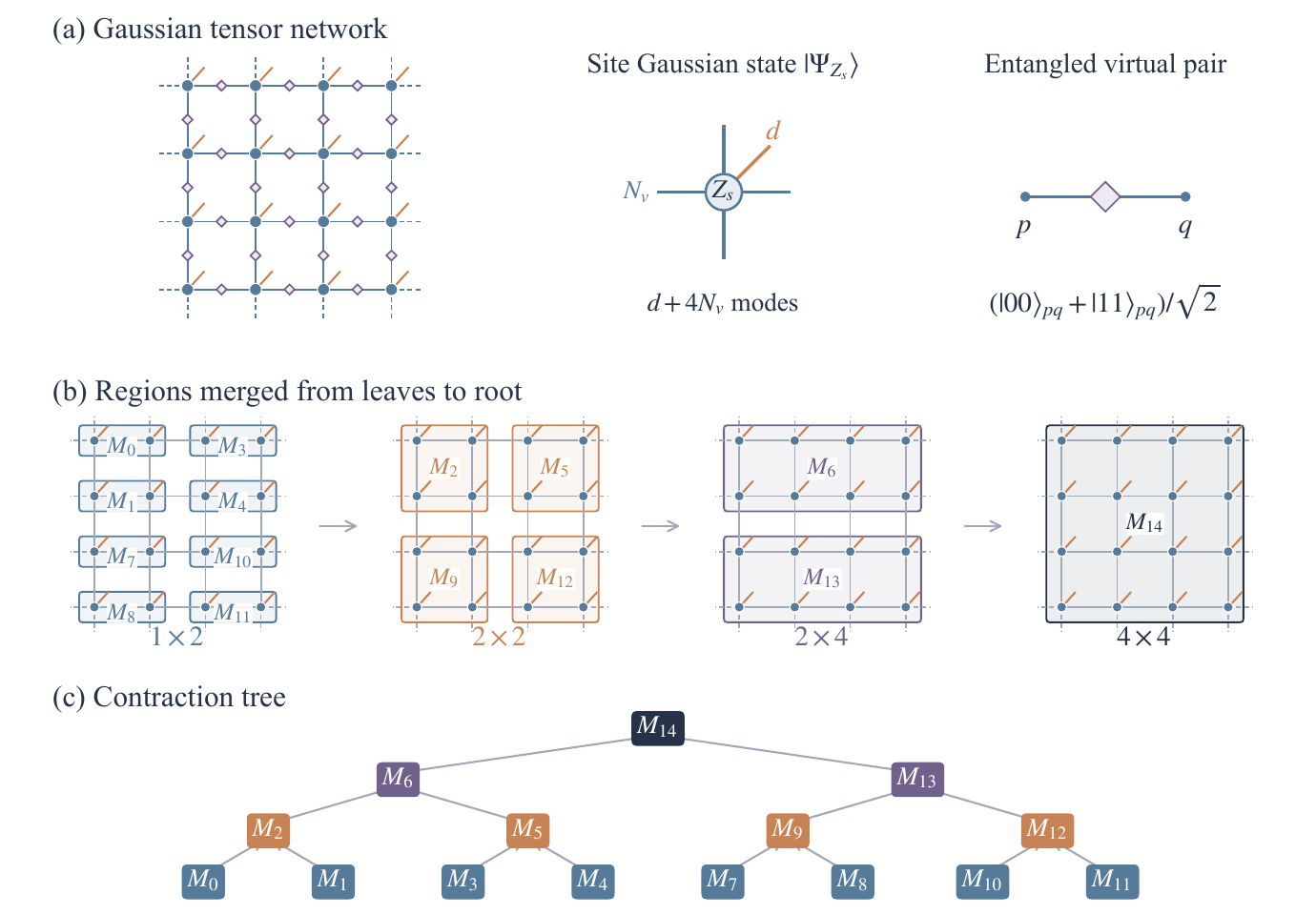}
    \caption{Gaussian tensor network and its hierarchical contraction on a $4\times4$ periodic lattice. (a) Network geometry with blue site tensors, purple entangled virtual pairs and orange physical legs. Dashed legs indicate periodic continuations. (b) Neighboring regions merge from two-site leaves into $2\times2$ blocks, $2\times4$ strips and the full lattice. (c) Contraction tree.}
    \label{fig:contraction_schematic}
\end{figure}

A Gaussian tensor-network state is obtained from a product of local Gaussian states by projecting virtual-mode pairs onto maximally entangled states~\cite{Kraus_PRA10_fPEPS}. As illustrated in Fig.~\ref{fig:contraction_schematic}(a), it can be written as
\begin{equation}
    \ket{\Psi}=(\mathbf{1}_{\mathrm{phys}}\otimes\bra{\Omega})\ket{G},
    \label{eq:network_state}
\end{equation}
where $\ket{G}$ is the product of $N$ local Gaussian states, each parametrized by a skew-symmetric matrix $Z_s$,
\begin{equation}
    \ket{G}=\bigotimes_{s=1}^{N}\ket{\Psi_{Z_s}}
    =\ket{\Psi_Z},
    \qquad Z=\bigoplus_{s=1}^{N}Z_s.
    \label{eq:network_direct_sum}
\end{equation}
Each site has $d$ physical modes and $N_v$ virtual modes on each of four legs, so $Z_s$ has dimension $d+4N_v$, with $d=2$ for the Hubbard model. The bond state $\ket{\Omega}$ is a product of maximally entangled virtual pairs $(\ket{00}_{pq}+\ket{11}_{pq})/\sqrt{2}$. Its bra can be written as
\begin{equation}
    \bra{\Omega}
    =2^{-NN_v}\left(\bigotimes_{(p,q)\in \mathrm{bond}}\bra{11}_{pq}\right)
      \exp\!\left(\sum_{a<b}B_{ab}c_a^\dagger c_b^\dagger\right).
    \label{eq:network_bonds}
\end{equation}
The matrix $B$ has the same dimension $N(d+4N_v)$ as $Z$. Its only nonzero entries are $B_{pq}=-B_{qp}=1$ on each oriented virtual bond $(p,q)$. The amplitude of a physical configuration $n_I$ is
\begin{equation}
    \psi(n_I)=\braket{n_I}{\Psi}
    =(\bra{n_I}\otimes\bra{\Omega})\ket{G}.
    \label{eq:network_overlap}
\end{equation}
Substituting Eqs.~\eqref{eq:network_direct_sum} and~\eqref{eq:network_bonds} into Eq.~\eqref{eq:network_overlap} and applying Eq.~\eqref{eq:minor} gives
\begin{equation}
    \psi(n_I)=2^{-NN_v}\braket{\tilde n_I}{\Psi_{Z+B}}
    =2^{-NN_v}\Pf\!\left[\tilde n_I\star(Z+B)\star\tilde n_I\right].
    \label{eq:network_amplitude}
\end{equation}
Here $\tilde n_I$ denotes the configuration with physical occupations $n_I$ and all virtual modes occupied.
Reordering physical and virtual modes can introduce a fermionic sign. We use site-major ordering, with $(\mathrm{up},\mathrm{left},\mathrm{phys},\mathrm{down},\mathrm{right})$ within each site, so moving the $n_s$ occupied physical modes past two fully occupied virtual legs gives $(-1)^{2N_v n_s}=+1$.

The amplitude is obtained by computing the Pfaffian of the sparse matrix $M=\tilde n_I\star(Z+B)\star\tilde n_I$. Following a strategy similar to that in Ref.~\cite{Thomas_PRE13_Pfaffian}, we evaluate this Pfaffian by eliminating modes in pairs from the leaves to the root of the nested-dissection tree in Fig.~\ref{fig:contraction_schematic}(b,c).

Each tree node corresponds to a lattice region, whose mode indices are divided into $C$, eligible for local elimination, and $U$, retained for the parent. Eliminating $C$ from the regional matrix $M$ yields a reduced matrix $M_{UU}^{\mathrm{new}}$ on $U$. With modes ordered as $(C,U)$ and $M_{CC}$ invertible, the Pfaffian Schur identity gives
\begin{equation}
    M_{UU}^{\mathrm{new}}=M_{UU}-M_{UC}M_{CC}^{-1}M_{CU},
    \qquad
    \Pf M=\Pf M_{CC}\,\Pf M_{UU}^{\mathrm{new}}.
    \label{eq:front_schur}
\end{equation}
For a pair $k,r\in C$ with pivot $a=M_{kr}\ne0$, Eq.~\eqref{eq:front_schur} reduces to
\begin{equation}
    \begin{aligned}
        M'_{ij}&=M_{ij}-\frac{M_{ir}M_{kj}-M_{ik}M_{rj}}{a},
        \qquad i,j\in(C\cup U)\setminus\{k,r\},\\
        \Pf M&=\operatorname{sgn}(\pi_{kr})\,a\,\Pf M'.
    \end{aligned}
    \label{eq:front_pair_update}
\end{equation}
\begin{samepage}
Here $\pi_{kr}$ brings $(k,r)$ first in the mode ordering and preserves the order of the remaining modes. To improve numerical stability, we select the largest-magnitude entry within $C$ for each candidate row and accept a nonzero pivot only if its magnitude is at least $u$ times the maximum magnitude in the full row over $C\cup U$. During elimination, we record the original mode labels in the ordered list $\mathcal P$. The full procedure is summarized in Algorithm~\ref{alg:front}.
\par
\end{samepage}

\begin{algorithm}
\caption{Local Pfaffian elimination with threshold pivoting.}
\label{alg:front}
\begin{alglines}
\item \textbf{Input:} $M$, index sets $C$ and $U$, threshold $0<u\leq1$.
\item $\mathcal P\gets()$, $\ell\gets0$, $\sigma\gets1$.
\item \textbf{while} $|C|\geq2$ \textbf{do}
\item \algind{1} Find the first $k\in C$ for which
\item \algind{2} $r=\arg\max_{j\in C\setminus\{k\}}|M_{kj}|,\qquad a=M_{kr}\ne0,$
\item \algind{2} $|a|\geq u\max_{j\in C\cup U}|M_{kj}|$.
\item \algind{1} \textbf{if} no such $k$ exists \textbf{then break}.
\item \algind{1} $\ell\gets\ell+\log|a|$, $\sigma\gets\sigma\operatorname{sgn}(a)$.
\item \algind{1} Append the original mode labels $(k,r)$ to $\mathcal P$.
\item \algind{1} Apply the matrix update in Eq.~\eqref{eq:front_pair_update} to $M$ and remove $k,r$ from $C$.
\item \textbf{end while}
\item \textbf{Return:} $\ell,\sigma,\mathcal P,M,C$.
\end{alglines}
\end{algorithm}

\begin{samepage}
If no admissible pair remains, modes left in $C$ are delayed to the parent, as in delayed pivoting for sparse multifrontal factorization~\cite{Duff_TOMS83_Multifrontal,Duff_SIAM05_Pivoting}. The parent then combines its children's remaining matrices with its assigned entries and repeats Algorithm~\ref{alg:front}. Once all modes are eliminated, 
$\psi(n_I) = 2^{-NN_v}\,\mathrm{sgn}(\pi)\prod_{t=1}^{K} a_t$,
which we evaluate in signed-log form:
\begin{equation}
    \begin{aligned}
        \operatorname{sgn}\psi(n_I)&=\operatorname{sgn}(\pi)\prod_{t=1}^{K}\operatorname{sgn}(a_t),\\
        \log|\psi(n_I)|&=\sum_{t=1}^{K}\log|a_t|-NN_v\log2.
    \end{aligned}
    \label{eq:pf_product}
\end{equation}
\end{samepage}
Here $K$ counts the eliminated pairs, and $\pi$ maps the selected site-major order to $\mathcal P=(k_1,r_1,\ldots,k_K,r_K)$. Pivot signs and log magnitudes accumulate during elimination, and the permutation parity is applied at the end. With nested-dissection ordering and bounded pivot delays, the contraction requires $O(N_v^3N^{3/2})$ work and $O(N_v^2N\log N)$ storage~\cite{George_SIAM73_NestedDissection,Thomas_PRE13_Pfaffian}.

\section{Neural-network architecture}
\label{app:network}

\begin{figure}[!htbp]
 \centering
 \includegraphics[width=0.96\textwidth]{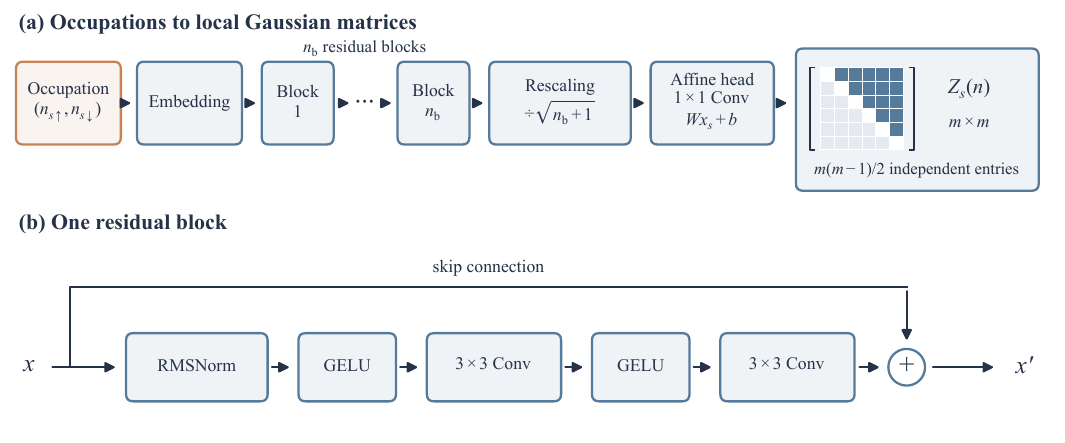}
 \caption{Residual CNN for NN-GPEPS. (a) Occupation embedding, residual blocks, rescaling and an affine head generate the Gaussian site matrices. Blue cells mark the independent matrix entries. (b) Pre-activation residual block.}
 \label{fig:network-architecture}
\end{figure}

\begin{table}[!htbp]
 \caption{Residual CNN settings for the $N_v=3$ and $4$ ansatzes. Parameter totals include the embedding and output head.}
 \label{tab:network-architecture}
 \begin{ruledtabular}
 \begin{tabular}{lcc}
  & $N_v=3$ & $N_v=4$ \\
  \colrule
  Local Gaussian matrix size $m$ & $14$ & $18$ \\
  Independent outputs per site & $91$ & $153$ \\
  Feature channels $C$ & $91$ & $153$ \\
  Residual blocks $n_{\rm b}$ & $16$ & $24$ \\
  Trainable parameters & $2\,396\,485$ & $10\,144\,053$ \\
 \end{tabular}
 \end{ruledtabular}
\end{table}

We generate the Gaussian site matrices with a real-valued residual CNN~\cite{He_CVPR16_ResNet} (Fig.~\ref{fig:network-architecture}). A learned embedding maps each local occupation state to a $C$-dimensional feature vector, with the same embedding table used at every site. These features then pass through a sequence of $n_{\rm b}$ pre-activation residual blocks~\cite{He_ECCV16_ResNetIdentity} built from $3\times3$ convolutions with circular padding. Each block applies a parameter-free RMSNorm~\cite{Zhang_NIPS19_RMSNorm} over all channels and sites, followed by GELU nonlinearities~\cite{Hendrycks_arxiv23_GELU} before each convolution [Fig.~\ref{fig:network-architecture}(b)].

The features are rescaled by $1/\sqrt{n_{\rm b}+1}$ before an affine $1\times1$ head maps them to the strict upper triangle of $Z_s(n;\theta)$. We keep the channel width equal to the output dimension, $C=m(m-1)/2$, where $m=d+4N_v$ and $d=2$. Antisymmetry determines the remaining entries. The network settings are listed in Table~\ref{tab:network-architecture}.

We initialize the embedding table with standard normal entries, use LeCun normal initialization~\cite{LeCun_NNTT98_EfficientBackProp} for convolution weights, and set all biases to zero. The last convolution of every residual block is additionally scaled by $0.1$ ($N_v=3$) or $0.02$ ($N_v=4$) at initialization, so that each block starts close to the identity map. All learned maps are shared across sites, so the parameter count is independent of lattice size and the same weights can be transferred between periodic lattices.

\section{Runtime benchmarks}
\label{app:jit}
\label{app:benchmark}

\subsection{CNN and contraction costs}

NN-GPEPS amplitude evaluation consists of constructing a configuration-dependent Gaussian tensor network with the CNN, followed by its contraction. We benchmark the CNN evaluation in TF32, Gaussian tensor network contraction in FP32, and the complete amplitude evaluation on A100-SXM4-80GB GPUs of NERSC Perlmutter. These measurements use the $N_v=3$ network of Appendix~\ref{app:network}, with identical initial weights across lattice sizes and randomly sampled half-filled configurations, without symmetry projection.

We report GPU-synchronized timings per amplitude after compilation and warm-up, using the batch sizes listed in Table~\ref{tab:forward_cost}.

\begin{figure}[!htbp]
 \centering
 \includegraphics[width=4.1in]{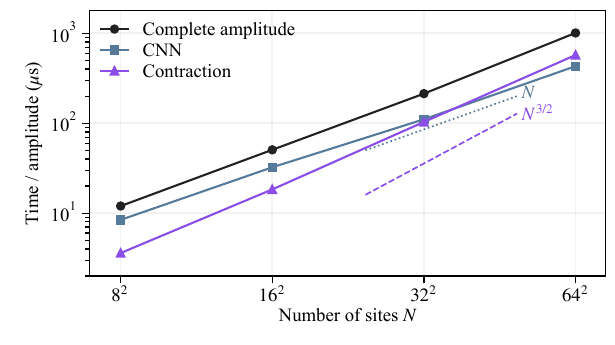}
 \caption{Time per amplitude for the complete evaluation, CNN and contraction of the $N_v=3$ network on one A100-SXM4-80GB GPU of NERSC Perlmutter. Points show median timings. Dotted and dashed guides indicate $N$ and $N^{3/2}$, respectively.}
 \label{fig:forward_cost}
\end{figure}

Figure~\ref{fig:forward_cost} and Table~\ref{tab:forward_cost} summarize the results. The ratio of CNN time to complete evaluation time decreases from $70\%$ at $8\times8$ to $43\%$ at $64\times64$. At the latter size, contraction accounts for the larger contribution and the complete evaluation requires $1.009$\,ms per amplitude.  The contraction times are plotted in Fig.~\ref{fig:scaling}.

\begin{table}[!htbp]
\caption{CNN, contraction and complete amplitude-evaluation times in microseconds, with batch size $B$, for the $N_v=3$ network on a Perlmutter A100 80GB GPU, as shown in Fig.~\ref{fig:forward_cost}. The final column gives the CNN time divided by the complete evaluation time. The CNN runs in TF32 and the contraction in FP32.}
\label{tab:forward_cost}
\begin{ruledtabular}
\small\begin{tabular}{crrrrr}
Lattice & $B$ & CNN & Contraction & Complete & CNN share \\
\colrule
$8\times8$ & $16384$ & $8.42$ & $3.60$ & $11.99$ & $70\%$ \\
$16\times16$ & $2048$ & $32.35$ & $18.37$ & $50.62$ & $64\%$ \\
$32\times32$ & $1024$ & $110.56$ & $103.29$ & $213.21$ & $52\%$ \\
$64\times64$ & $512$ & $430.76$ & $573.16$ & $1009.17$ & $43\%$ \\
\end{tabular}
\end{ruledtabular}
\end{table}

Table~\ref{tab:throughput_model} reports the floating-point operation count $F$ per contraction, the achieved throughput $F/t$ and the model FLOPs utilization $\mathrm{MFU}=F/(tP)$, where $P$ = 19.5 TFLOP/s is the A100's peak FP32 throughput.

\begin{table}[!htbp]
\centering
\begin{minipage}{0.96\textwidth}
\centering
\caption{Floating-point operation counts, times per amplitude, achieved throughput and model FLOPs utilization (MFU) of Gaussian tensor network contraction on one Perlmutter A100 80GB GPU, at $N_v=3$ in FP32 with the batch sizes of Table~\ref{tab:forward_cost}.}
\label{tab:throughput_model}
{\small
\begin{tabular*}{\linewidth}{@{\extracolsep{\fill}}lrrrr@{}}
\hline\hline
Lattice & $F$ (MFLOP) & Time ($\mu$s) & $F/t$ (TFLOP/s) & MFU \\
\hline
$8\times8$ & 1.04 & 3.60 & 0.29 & 1.5\% \\
$16\times16$ & 6.71 & 18.37 & 0.37 & 1.9\% \\
$32\times32$ & 47.4 & 103.29 & 0.46 & 2.4\% \\
$64\times64$ & 353.5 & 573.16 & 0.62 & 3.2\% \\
\hline\hline
\end{tabular*}}
\end{minipage}
\end{table}

\subsection{Single- and double-precision contraction}

\begin{figure}[!htbp]
 \centering
 \includegraphics[width=5.2in]{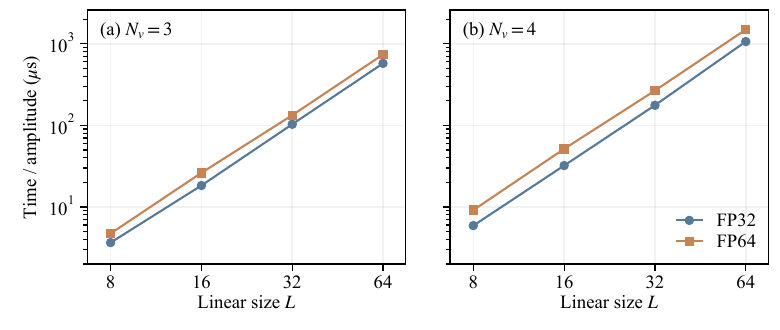}
 \caption{FP32 and FP64 Gaussian tensor network contraction times on one A100-SXM4-80GB GPU of NERSC Perlmutter for (a) $N_v=3$ and (b) $N_v=4$. Each precision pair uses identical inputs and batch sizes. Points show median timings.}
 \label{fig:contraction-precision}
\end{figure}

We next compare FP32 and FP64 Gaussian tensor network contraction times at $N_v=3$ and $4$ on the same GPU. To isolate the contraction cost, the site-matrix entries are generated directly with standard deviation $0.5$ and rounded to FP32 before conversion to either working precision. Both calculations therefore use identical matrix entries and half-filled configurations. All contraction benchmarks use two-site leaves, a pivot threshold of $0.05$ and $20$ delay slots per child. In the FP32 implementation, global-front updates use FP64 Tensor Cores while the contraction matrices remain stored in FP32. FP32 and FP64 timings use identical inputs and batch sizes and include only the Gaussian tensor-network contraction. Figure~\ref{fig:contraction-precision} and Table~\ref{tab:contraction-precision} show that FP32 is faster throughout the tested range.

\begin{table}[!htbp]
\caption{Signed-amplitude contraction time in microseconds on one A100-SXM4-80GB GPU of NERSC Perlmutter, with batch size $B$.}
\label{tab:contraction-precision}
\begin{ruledtabular}
\small\begin{tabular}{crrrrrrrr}
 & \multicolumn{4}{c}{$N_v=3$} & \multicolumn{4}{c}{$N_v=4$} \\
Lattice & $B$ & FP32 & FP64 & FP64/FP32 & $B$ & FP32 & FP64 & FP64/FP32 \\
\colrule
$8\times8$ & $8192$ & $3.65$ & $4.73$ & $1.29$ & $8192$ & $5.91$ & $9.15$ & $1.55$ \\
$16\times16$ & $2048$ & $18.32$ & $26.19$ & $1.43$ & $2048$ & $32.16$ & $51.51$ & $1.60$ \\
$32\times32$ & $1024$ & $103.12$ & $133.03$ & $1.29$ & $1024$ & $176.20$ & $266.74$ & $1.51$ \\
$64\times64$ & $512$ & $574.91$ & $739.29$ & $1.29$ & $512$ & $1064.17$ & $1493.48$ & $1.40$ \\
\end{tabular}
\end{ruledtabular}
\end{table}

\section{Optimization and measurements}
\label{app:recipe}

\subsection{Sampling and optimization}

We use the mixed Metropolis sampler implemented in Quantax~\cite{quantax}, with equal probabilities for nearest-neighbor electron hops and site exchanges. Each update uses $8192$ configurations in a fixed $(N_\uparrow,N_\downarrow)$ sector, separated by $4N$ proposals per chain. Sampling follows the squared amplitude of the wavefunction being optimized.

We optimize the network with the Quantax implementation of AdamSR~\cite{quantax}, with first- and second-moment decay factors $\mu=0.95$ and $\beta=0.995$, respectively. The associated least-squares problems are solved using an adaptation of Quantax's LSMR solver~\cite{quantax,Fong_SIAM11_LSMR} with relative tolerance $10^{-4}$, absolute tolerance $0$ and at most $500$ iterations, without an explicit diagonal shift. Local energies entering the gradient are clipped to the sampled mean $\pm5\sigma$, where $\sigma$ is their sampled standard deviation~\cite{Gu_NC26_NQS-Hubbard}.

We constrain the Euclidean norm of each parameter update~\cite{Gu_NC26_NQS-Hubbard}. At iteration $t$, the proposed update $\Delta\theta_t$ is rescaled as
\begin{equation}
 \Delta\theta_t \leftarrow
 \frac{C_t}{\max(C_t,\|\Delta\theta_t\|_2)}\,\Delta\theta_t,
 \qquad
 C_t=\frac{C_0}{1+\max(t-t_0,0)/T}.
 \label{eq:update-norm-bound}
\end{equation}
The norm bound remains at $C_0$ until iteration $t_0$, then decreases on the scale $T$. Updates whose norms lie below this bound are unchanged. A typical setting is $C_0=0.5$, $t_0=1000$ and $T=1500$, with tighter norm bounds and adjusted schedules for continued and projected optimization.

\subsection{Training and measurements}

We first optimize without symmetry projection (U), then continue with the eight-image projection. Table~\ref{tab:recipe} gives the training lengths and contraction settings, using the architectures of Appendix~\ref{app:network}. CNN products use TF32 during training.

\begin{table}[!htbp]
 \caption{Training and contraction settings at $U/t=8$ and $1/8$ doping. U and P denote unprojected and symmetry-projected optimization, respectively. Training steps are listed as U + P, the $16\times16$ states start from weights trained on $16\times4$.}
 \label{tab:recipe}
 \begin{ruledtabular}
 \begin{tabular}{lcccc}
  Lattice & \multicolumn{2}{c}{$16\times4$} & \multicolumn{2}{c}{$16\times16$} \\
  $N_v$ & $3$ & $4$ & $3$ & $4$ \\
  \colrule
  Electrons $(N_\uparrow,N_\downarrow)$ & $(28,28)$ & $(28,28)$ & $(112,112)$ & $(112,112)$ \\
  Proposals between samples & $256$ & $256$ & $1024$ & $1024$ \\
  Gaussian tensor network contraction precision & FP32 & FP64 & FP32 & FP64 \\
  Pivot threshold & $0.05$ & $0.05$ & $0.05$ & $0.1/0.05$ \\
  Delay slots per child & $20$ & $16$ & $20$ & $24/20$ \\
  Training steps (U + P) & $10000+5000$ & $10000+5000$ & $4321+1520$ & $5340+1500$ \\
 \end{tabular}
 \end{ruledtabular}
\end{table}

When no pivot passes the relative threshold of Algorithm~\ref{alg:front}, the corresponding modes are delayed and eliminated at the parent node. Because the GPU implementation uses fixed-size arrays, each parent node reserves a fixed number of delay slots for each child; we refer to this number as the delay capacity.
A sampled configuration is dropped from the energy and gradient estimates of an update when one of its connected configurations exceeds the delay capacity.
All reported energies and correlations are measured at fixed neural parameters with FP32 CNN products and FP32 Gaussian tensor network contraction for both sampling and observable evaluation. Table~\ref{tab:measurement} lists the measurement settings. A configuration whose contraction exceeds the delay capacity is assigned zero amplitude, both when proposed by the sampler and when it enters a local estimator, so that sampling and evaluation refer to the same state and the measured energy is its variational energy. In the reported measurements no configuration exceeded the capacity. Statistical uncertainties are the largest block-jackknife standard errors obtained with blocks of $1$, $2$, $4$ and $8$ consecutive measurement batches.

\begin{table}[!htbp]
 \caption{Measurement settings for the projected states of Table~\ref{tab:recipe} and for the $16\times16$ states transferred to $32\times32$ (Appendix~\ref{app:precision}), the latter measured without (U) or with (P) the projection. $N$ is the number of sites.}
 \label{tab:measurement}
 \begin{ruledtabular}
 \begin{tabular}{lcccccc}
  Lattice & \multicolumn{2}{c}{$16\times4$} & \multicolumn{2}{c}{$16\times16$} & \multicolumn{2}{c}{$32\times32$} \\
  $N_v$ & $3$ & $4$ & $3$ & $4$ & $3$ & $4$ \\
  \colrule
  CNN products & FP32 & FP32 & FP32 & FP32 & FP32 & FP32 \\
  Gaussian tensor network contraction precision & FP32 & FP32 & FP32 & FP32 & FP32 & FP32 \\
  Pivot threshold & $0.05$ & $0.05$ & $0.05$ & $0.05$ & $0.05$ & $0.05$ \\
  Delay slots per child & $24$ & $20$ & $32$ & $32$ & $64$ & $64$ \\
  Equilibration proposals per chain & $40N$ & $40N$ & $40N$ & $40N$ & $40N$ & $40N$ \\
  Proposals between batches & $4N$ & $4N$ & $4N$ & $4N$ & $4N$ & $4N$ \\
  Measurement batches & $64$ & $64$ & $64$ & $64$ & $64$ (U), $32$ (P) & $64$ (U), $32$ (P) \\
  Configurations per batch & $8192$ & $8192$ & $8192$ & $8192$ & $1024$ & $1024$ \\
 \end{tabular}
 \end{ruledtabular}
\end{table}

\section{Evaluation after weight transfer}
\label{app:precision}

We transfer the weights of the unprojected $N_v=3$ and $4$ networks trained on $16\times16$ with TF32 CNN products to $32\times32$ and measure their energies and local-energy variances. We compare three schemes: FP32 sampling and evaluation, TF32 sampling with FP32 reweighting, and TF32 sampling and evaluation. All three use FP32 Gaussian tensor network contraction.

Table~\ref{tab:fixed-reweight-pending} shows similar energies across the three schemes, but TF32 evaluation increases the variance relative to FP32 by a factor of about $2.3$ for $N_v=3$ and $4$. FP32 reweighting agrees with direct FP32 sampling in both energy and variance.

Let $\psi(n)$ and $\widetilde\psi(n)$ denote amplitudes evaluated with FP32 and TF32 CNN products. Direct sampling from $p(n)\propto|\psi(n)|^2$ gives the usual mean of $E_{\rm loc}^{\rm FP32}$. Samples from $q(n)\propto|\widetilde\psi(n)|^2$ instead receive importance weights
\begin{equation}
 r_i=\frac{|\psi(n_i)|^2}{|\widetilde\psi(n_i)|^2},\qquad
 w_i=\frac{r_i}{\sum_j r_j}.
 \label{eq:precision-weights}
\end{equation}
The sums run over one measurement batch. Averaging the normalized moments over batches gives
\begin{equation}
\begin{aligned}
 \widehat E_{\rm rw}&=\overline{\sum_i w_i E_{\rm loc}^{\rm FP32}(n_i)},\\
 \widehat{\sigma^2}_{\rm rw}&=\overline{\sum_i w_i\left|E_{\rm loc}^{\rm FP32}(n_i)\right|^2}-\left|\widehat E_{\rm rw}\right|^2.
\end{aligned}
 \label{eq:precision-reweight}
\end{equation}
Here the overbar denotes an equal-weight average over batches, and all local-energy ratios use FP32 amplitudes.

At each $N_v$, all schemes use the same source weights, held fixed throughout, and collect $64$ measurement batches of $1024$ configurations, with $4N$ Metropolis proposals per chain between batches. Each run starts from random configurations and equilibrates for $40N$ proposals per chain. The FP32 scheme is a separate run, while the two TF32-sampling estimators are evaluated on the same TF32-sampled configurations of one run.

Statistical uncertainties are the largest block-jackknife standard errors obtained with blocks of $1$, $2$, $4$ and $8$ consecutive measurement batches.

\begin{table}[!htbp]
\caption{Precision comparison for unprojected $N_v=3$ and $4$ states transferred from $16\times16$ to $32\times32$. Energies and local-energy variances are per site. All three schemes use FP32 Gaussian tensor network contraction and were run on NVIDIA A100 GPUs. Parentheses give statistical uncertainties.}
\label{tab:fixed-reweight-pending}
\begin{ruledtabular}
\begin{tabular}{lcc}
Sampling and estimator & $E/N$ & $\sigma^2/N$ \\
\colrule
\multicolumn{3}{c}{$N_v=3$} \\
\colrule
FP32 sampling and evaluation & $-0.757451(21)$ & $0.02407(51)$ \\
TF32 sampling, FP32 reweighting & $-0.757429(27)$ & $0.02398(76)$ \\
TF32 sampling and evaluation & $-0.757396(36)$ & $0.05573(81)$ \\
\colrule
\multicolumn{3}{c}{$N_v=4$} \\
\colrule
FP32 sampling and evaluation & $-0.758019(22)$ & $0.02013(42)$ \\
TF32 sampling, FP32 reweighting & $-0.758021(18)$ & $0.02051(68)$ \\
TF32 sampling and evaluation & $-0.757979(28)$ & $0.04781(75)$ \\
\end{tabular}
\end{ruledtabular}
\end{table}

Numerical errors in amplitude ratios can partly cancel in the mean energy while contributing strongly to its variance. The reported transfer results therefore use FP32 sampling and evaluation. The same importance weights can also enter the means and covariances in SR, allowing TF32 sampling with FP32 local energies during training, although we did not use that here.

\FloatBarrier
\section{Extrapolation}
\label{app:extrapolation}

We jointly extrapolate the symmetry-projected $N_v=3$ and $4$ training trajectories on each lattice to zero variance. 
Each fit uses a window of consecutive updates ending at the last update of
each projected trajectory: the final 100, 200 or 400 updates for $N_v=3$,
and the portion starting at 30, 40, 50, 60 or 70\% of the trajectory for
$N_v=4$. Within each window, we average the training estimates of $E/N$
and $\sigma^2/N$ over bins of consecutive updates, aligned to the end of
the trajectory.
Each line in Fig.~\ref{fig:extrapolation-summary}(a,b) combines all bins from one $N_v=3$ window and one $N_v=4$ window, using the same bin size for both. We fit these points to
\begin{equation}
E/N=\epsilon_0+a\,\sigma^2/N
\label{eq:variance-extrapolation}
\end{equation}
by orthogonal-distance regression, using errors on both axes estimated within each bin.

We repeat the fit using different training windows and bin sizes (Table~\ref{tab:extrapolation-summary}). The reported energy is the mean of the fitted intercepts. Its uncertainty covers the two-standard-error interval of every fit, giving
\begin{equation}
\begin{aligned}
16\times4:&\qquad E_0/N=-0.7673(5),\\
16\times16:&\qquad E_0/N=-0.7622(7).
\end{aligned}
\end{equation}
The energies and variances are the sampled training-window values, with TF32 CNN products. The quoted uncertainties reflect fitting sensitivity within the linear energy--variance model.

\begin{suppfigure}
\centering
\includegraphics[width=0.90\textwidth]{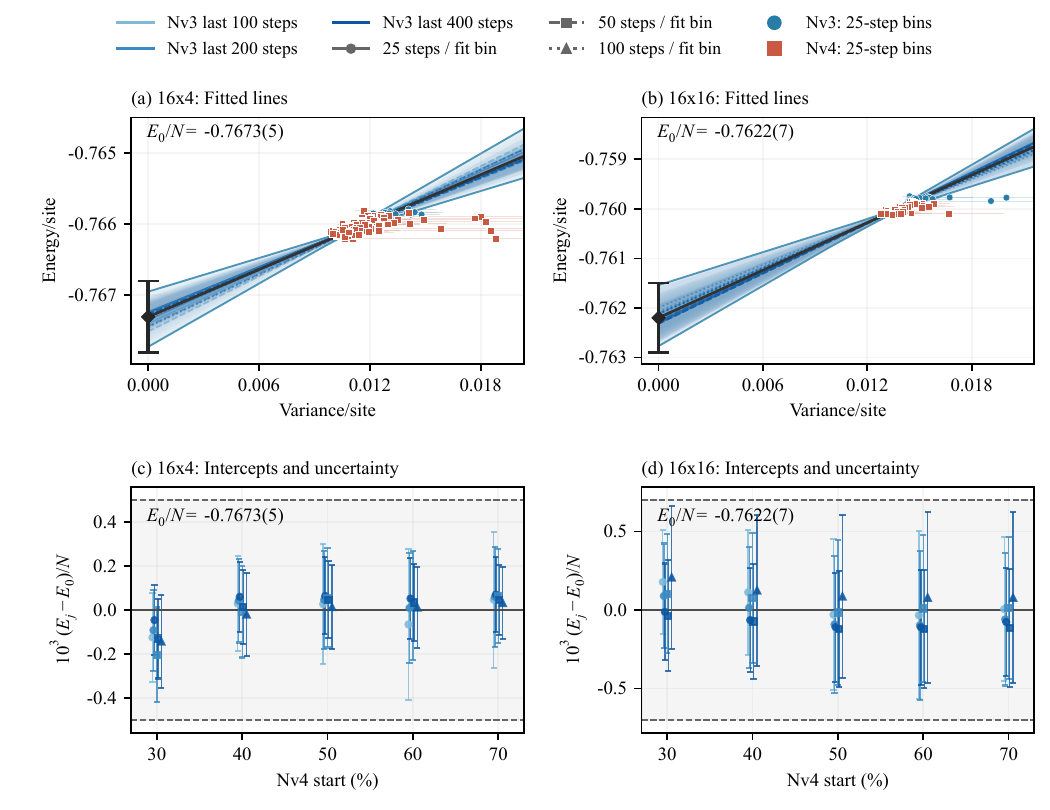}
\caption{Joint zero-variance extrapolation of the projected $N_v=3$ and $4$ trajectories. (a,b) Points use 25-update bins from the final 400 $N_v=3$ updates and the 30--100\% portion of the $N_v=4$ trajectory. Lines show fits using 25-, 50- and 100-update bins, with the dark line giving their mean fitted energy. The blue envelope contains their two-standard-error bands. (c,d) Fitted intercepts $E_j/N$ relative to their mean $E_0/N$ versus the $N_v=4$ window start, with two-standard-error bars. Black bars in (a,b) and gray bands in (c,d) mark the reported uncertainties.}
\label{fig:extrapolation-summary}
\end{suppfigure}

\begin{supptable}
\centering
\small
\caption{Joint zero-variance energies per site. Each row fixes the $N_v=3$ window length and bin size and summarizes five $N_v=4$ windows starting at 30, 40, 50, 60 and 70\% of the training trajectory. All windows end at the last recorded update and contain at least four complete bins. The energy range spans these five fits. $M$ counts retained/requested fits, and $s_{\max}$ is the largest fitting standard error of $E_0/N$.}
\label{tab:extrapolation-summary}
\begin{ruledtabular}
\begin{tabular}{lccrccc}
Lattice & $N_v=3$ final updates & Steps/bin & $M$ & Mean $E_0/N$ & Range of $E_0/N$ & $s_{\max}$ \\
\colrule
$16\times4$ & 100 & 25 & 5/5 & $-0.767322$ & $[-0.767429,-0.767258]$ & $0.000172$ \\
$16\times4$ & 200 & 25 & 5/5 & $-0.767292$ & $[-0.767397,-0.767245]$ & $0.000126$ \\
$16\times4$ & 200 & 50 & 5/5 & $-0.767319$ & $[-0.767509,-0.767240]$ & $0.000123$ \\
$16\times4$ & 400 & 25 & 5/5 & $-0.767265$ & $[-0.767351,-0.767235]$ & $0.000092$ \\
$16\times4$ & 400 & 50 & 5/5 & $-0.767303$ & $[-0.767435,-0.767258]$ & $0.000091$ \\
$16\times4$ & 400 & 100 & 5/5 & $-0.767326$ & $[-0.767448,-0.767273]$ & $0.000105$ \\
\colrule
$16\times16$ & 100 & 25 & 5/5 & $-0.762155$ & $[-0.762235,-0.762023]$ & $0.000267$ \\
$16\times16$ & 200 & 25 & 5/5 & $-0.762231$ & $[-0.762300,-0.762112]$ & $0.000236$ \\
$16\times16$ & 200 & 50 & 5/5 & $-0.762164$ & $[-0.762218,-0.762099]$ & $0.000234$ \\
$16\times16$ & 400 & 25 & 5/5 & $-0.762276$ & $[-0.762312,-0.762213]$ & $0.000183$ \\
$16\times16$ & 400 & 50 & 5/5 & $-0.762295$ & $[-0.762323,-0.762238]$ & $0.000189$ \\
$16\times16$ & 400 & 100 & 5/5 & $-0.762087$ & $[-0.762123,-0.761994]$ & $0.000272$ \\
\colrule
\multicolumn{4}{l}{Extrapolated energy, $16\times4$} & \multicolumn{3}{c}{$-0.7673(5)$} \\
\multicolumn{4}{l}{Extrapolated energy, $16\times16$} & \multicolumn{3}{c}{$-0.7622(7)$} \\
\end{tabular}
\end{ruledtabular}
\end{supptable}

\begin{samepage}
\section{Additional benchmarks and convergence}
\label{app:energy-benchmarks}
\label{app:observable-convergence}

\subsection{The $16\times4$ benchmark}
\label{app:strip-benchmark}

Table~\ref{tab:energy-16x4} compares energies and local-energy variances on the $16\times4$ lattice at $U/t=8$ and $1/8$ hole doping. The correlation functions below use the definitions given in the main text.

\begin{suppfigure}
\centering

\includegraphics[width=0.48\textwidth]{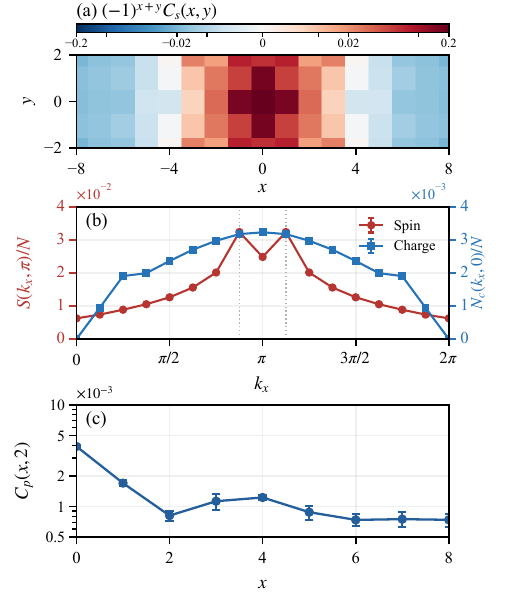}
\setlength{\abovecaptionskip}{4pt}
\caption{Correlations of the projected $16\times4$, $N_v=4$ NN-GPEPS state (Hubbard model $U/t=8$, $1/8$ doping). (a) Staggered spin correlations $(-1)^{x+y}C_s(x,y)$ at relative site separation $(x,y)$. The color scale is symmetric logarithmic and saturates outside $[-0.2,0.2]$. (b) Spin structure factor $S(k_x,\pi)/N$ (left axis) and charge structure factor $N_c(k_x,0)/N$ (right axis). (c) Pair correlation $C_p(x,2)$. Error bars indicate statistical uncertainties.}
\label{fig:orders-16x4}
\end{suppfigure}
\end{samepage}

\begin{supptable}
\centering

\caption{Energies and local-energy variances per site at $U/t=8$ and 1/8 hole doping on the periodic $16\times4$ lattice. Parentheses give reported uncertainties. The NN-GPEPS extrapolation procedure is described in Appendix~\ref{app:extrapolation}. A dash denotes an unavailable variance.}
\label{tab:energy-16x4}
\footnotesize\setlength{\tabcolsep}{3pt}
\begin{ruledtabular}
\begin{tabular}{lcc}
Method & $E/N$ & $\sigma^2/N$ \\
\colrule
Transformer backflow~\cite{Gu_NC26_NQS-Hubbard} & $-0.76298$ & --- \\
HFPS~\cite{Chen_PNAS26_HFPS} & $-0.76413(3)$ & $0.0313$ \\
ACE~\cite{Gu_PRX-Intelligence_SCALE} & $-0.76623(1)$ & --- \\
NNBF~\cite{Loehr_arxiv25_NNBF} & $-0.766073(6)$ & --- \\
NN-GPEPS, $N_v=3$ & $-0.765883(26)$ & $0.0133(7)$ \\
NN-GPEPS, $N_v=4$ & $-0.766152(16)$ & $0.01095(42)$ \\
NN-GPEPS (extrapolated) & $-0.7673(5)$ & --- \\
\end{tabular}
\end{ruledtabular}
\end{supptable}

Figure~\ref{fig:orders-16x4} shows the correlations of the projected $N_v=4$ state. The alternating spin correlations have a modulated envelope, and $S(k_x,\pi)$ peaks at $k_x=\pi\pm\pi/8$, identifying spin stripes with period 16. The charge structure factor has a weaker shoulder near $k_x=\pi/4$, consistent with the stripe pattern. Pair correlations decrease rapidly at short distances and give a far-distance average $\Delta_{\rm SC}^2=8.4(4)\times10^{-4}$. The stripe therefore displays pronounced spin-stripe correlations together with weak pairing at finite distances, qualitatively consistent with earlier HFPS results~\cite{Chen_PNAS26_HFPS}.

\par\noindent\begin{minipage}{\linewidth}
\subsection{Convergence with $N_v$}

We compare energies and observables of the projected $N_v=3$ and $4$ states at $U/t=8$ and $1/8$ hole doping. Table~\ref{tab:energy-source} compares training energies and independent FP32 measurements on the $16\times16$ lattice. The correlation measurements follow Appendix~\ref{app:recipe}. Since the neural networks also differ in width and depth (Appendix~\ref{app:network}), the comparison tests the achieved accuracy of the variational states rather than $N_v$ alone.

\begin{supptable}
\caption{Energies per site on $16\times16$, optimized and evaluated without or with the eight-image projection. Asterisks denote training-window means over the final $200$ ($N_v=3$, without projection), $50$ ($N_v=3$, with projection) or $100$ ($N_v=4$) updates, chosen where each trajectory had flattened. Independent fixed-parameter measurements use FP32 sampling and evaluation. Parentheses give statistical uncertainties.}
\label{tab:energy-source}
\begin{ruledtabular}
\begin{tabular}{clcc}
$N_v$ & Estimate & Without projection & With projection \\
\colrule
3 & Training-window mean & $-0.758558(8)^{*}$ & $-0.759839(13)^{*}$ \\
3 & Fixed-parameter FP32 measurement & & $-0.759835(14)$ \\
4 & Training-window mean & $-0.758920(13)^{*}$ & $-0.760102(8)^{*}$ \\
4 & Fixed-parameter FP32 measurement & & $-0.760115(11)$ \\
\end{tabular}
\end{ruledtabular}
\end{supptable}
\end{minipage}\par

Figure~\ref{fig:observable-convergence} shows the connected pair correlation of Eq.~\eqref{eq:pair-correlations} along the cut $(x,2)$ on both lattices. In Table~\ref{tab:observable-convergence}, $M_{\rm str}^2$ averages $S(\mathbf Q)/N$ over the two wavevectors $(\pi\pm\pi/8,\pi)$ on $16\times4$ and the four wavevectors $(\pi\pm\pi/8,\pi)$ and $(\pi,\pi\pm\pi/8)$ on $16\times16$. The pairing measure $\Delta_{\rm SC}^2$ averages over all displacements with shortest periodic distance $d\ge d_{\max}/2$, giving cutoffs $\sqrt{17}$ and $4\sqrt2$, respectively. Both order parameters $M_{\rm str}^2$ and $\Delta_{\rm SC}^2$ are consistent between the projected $N_v=3$ and $4$ states.

\begin{suppfigure}
\centering
\includegraphics[width=5.4in]{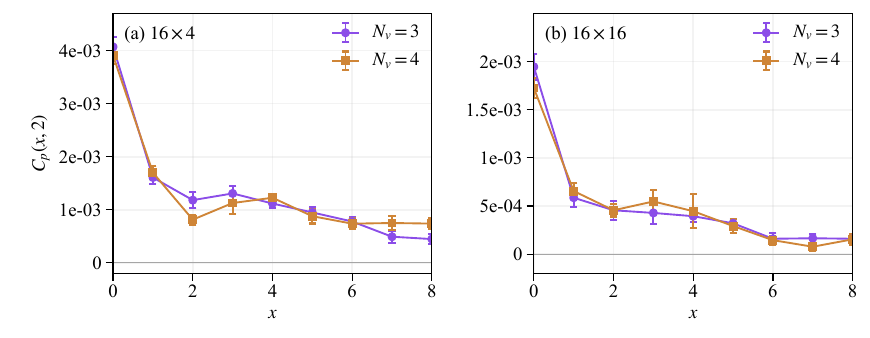}
\caption{Connected pair correlations $C_p(x,2)$ of projected NN-GPEPS states on (a) $16\times4$ and (b) $16\times16$. Error bars indicate statistical uncertainties.}
\label{fig:observable-convergence}
\end{suppfigure}

\begin{supptable}
\caption{Stripe and pairing measures of projected NN-GPEPS states. Parentheses give statistical uncertainties.}
\label{tab:observable-convergence}
\begin{ruledtabular}
\begin{tabular}{cccc}
Lattice & $N_v$ & $M_{\rm str}^2$ & $\Delta_{\rm SC}^2$ \\
\colrule
$16\times4$ & 3 & $0.032464(31)$ & $8.18(33)\times10^{-4}$ \\
 & 4 & $0.032403(27)$ & $8.43(41)\times10^{-4}$ \\
\colrule
$16\times16$ & 3 & $0.0138749(78)$ & $1.23(13)\times10^{-4}$ \\
 & 4 & $0.013725(10)$ & $1.37(20)\times10^{-4}$ \\
\end{tabular}
\end{ruledtabular}
\end{supptable}

\end{document}